\RequirePackage{fix-cm}

\documentclass[onecollarge]{svjour2}           % twocolumn
\smartqed 
\usepackage{graphicx}
\begin{document}

\title{A new analysis of the GJ581 extrasolar planetary system }

\author{M. Tadeu dos Santos \and G. G. Silva \and S. Ferraz-Mello \and T.A. Michtchenko} 

\institute{M.Tadeu dos Santos \at
              Instituto de Astronomia, Geof\'isica e Ci\^encias Atmosf\'ericas(IAG) \\ Universidade de S\~ao Paulo \\ S\~ao Paulo � Brasil \\
              \email{mtadeu@astro.iag.usp.br}           %  \\            
}
\date{Received: date / Accepted: date}

\maketitle

\begin{abstract}

We have done a new analysis of the available observations for the GJ581 exoplanetary system. Today this system is controversial due {to choices that can be done} in the orbital determination. The main ones are the ocurrence of aliases and the additional bodies - the planets f and g - announced in Vogt et al. 2010. Any dynamical study of exoplanets {requires the good knowledge} of the orbital elements and the investigations involving the planet g are particularly interesting, {since this body would lie in the} Habitable Zone (HZ) of the star GJ581. This region, for this system, is very attractive of the dynamical point of view due to several resonances {of two and three bodies present there.}
In this work, {we investigate the conditions under which the planet g may exist}. We stress the fact that the planet g is intimately related with the orbital elements of the planet d; more precisely, we conclude that it is not possible to disconnect its existence from the determination of the eccentricity of the planet d. Concerning the planet f, we have found one solution with period $\approx 450$ days, but we are judicious about any affirmation concernig this body because its signal is in the threshold of detection and the high period is in a spectral region where the ocorruence of aliases is very common. Besides, we {outline some dynamical features} of the habitable zone with the dynamical map and point out the role played by some resonances laying there.     


\end{abstract}

\section{Introduction}
%
%\label{intro}
%
The diversity of the exoplanets known today is quite big: large and small masses, semi-major axes, eccentricities, etc. However there are only very few planets known in the regions where it is possible to find favourable conditions for the development of life, the habitable zones. The importance of the planetary system around the {0.31 $M_{\odot}$}, M3V star GJ581 {is in this context:} According to von Paris et al. (2011) and Kaltenegger et al.(2011), GJ581 d is a {potentially} habitable exoplanet. Besides, a fifth putative planet, GJ581 g,  could exist in the central part of the habitable zone. This gives particular importance to the study of the dynamics and stability of this system. The possibility of a near-resonant scenario makes necessary a good knowledge of the masses and elements. Dynamical studies are often very time-consuming and near resonances critically depend on some elements. It is thus necessary to have an extended analysis of the existing observations so that we may know in advance the level of confidence that can be attributed to the masses and elements.
\\
This analysis was done using the 119 HARPS radial velocity measurements of this system available in the VizieR data base and the 109 Lick observations done between {August, 1999} and May, 2010. From the first set, Mayor et al (2009; hereafter M09), concluded the existence of four planets, while from the second one, Vogt et al (2010; hereafter V10), claimed the discovery of two additional bodies: GJ581 f, with period $\approx 433$ days and GJ581 g with period $\approx 36$ days.
\\
We have done a new analysis of these observations and we have investigated some of the problems pointed in earlier works. One of them, very common in the analysis of time series, is the aliasing indetermination of the period of GJ581 d. As shown by Fabrycky $\&$ Dawson {(2010)}, there are two almost equivalent solutions for planet d, one of them with period $\approx 67$ days, as found in M09 and, the other, $\approx 1.0124$ days. Unfortunately, this problem is intimately related with the spacing of the data in the observations time series and there is no mathematical way to distinguish which one is the correct period. However the joint analysis of the two time series allowed us to decide in favour of the longest period among the two possible ones. The other problem is the existence of the two planets claimed in V10. The planet GJ581 f is technically detectable and the second - GJ581 g - is also possible if the orbits of the other planets in the system (mainly GJ581 d) are assumed to be circular.
\\
The orbit determinations resulting from the two existing data series and the problems related to the period of GJ581 d, are discussed in Section 2. Section 3 is devoted to the joint analysis of the two data sets and to the tests concerning the detection of GJ581 g. It also includes an analysis of the dynamics inside the Habitable Zone. The detection of the planet GJ581 f is discussed in Section 4, and, at last, Section 5 presents a final discussion and the conclusions.

\section{The previous orbit determination}

This section presents independent analyses based on the two existing series. In all cases, the optimization procedure used was the genetic algorithm proposed by Charbonneau {(1995, 2002)} followed by a simplex optimization (Press et al. 1992), It was done using the same package used in the study of the CoRoT 7 system (Ferraz-Mello et al. 2011). A basic tool in the analysis of the data was the DCDFT (date- compensated discrete Fourier transform; Ferraz-Mello, 1981), which was used as a diagnostic tool to assess the periodicities in the data. However, it is worth stressing the fact that in all models studied in this paper, the given periods result from the joint best-fit of all elements. The DCDFT spectra were mainly used to check if a given periodicity on the system can be obtained by random data showing the same distribution. These random samples were obtained by shuffling the data (but keeping the dates unchanged). The spectra obtained from the randomly distributed samples allow us to determine in each case a minimum level of confidence for the peaks. In order to be accepted, the peaks of the analyzed data must appear significantly higher than this minimum confidence level. In general, to define the confidence level, we have used 1000 random samples.
\\
For the sake of comparing results obtained with the different models analyzed in this paper, we adopt as goodness-of-fit estimator the weighted root mean square of the residuals (see Beaug\'e et al. 2008):
\begin{equation}
(wrms)^2 = \frac{S}{N-1} \sum_{i=1}^N \left( \frac{\left( O_i-C_i \right) ^2}{ \sigma_i^2 } \right)
\end{equation}
\\
where 
\\
\begin{equation}
S^{-1} = \frac{1}{N} \sum_{i=1}^N \left( \frac{1}{ \sigma_i^2 } \right).
\end{equation}
\\
{$O_i$ and $\sigma_i$ are the radial velocities measured and the errors, respectively. N is the number of observations, M the total number of parameters determined and $C_i$ the radial velocities calculated with the model considered.}
We also give, to ease comparisons with results in other papers, the quantity Q given by
\begin{equation}
Q = \frac{S}{N-M} \sum_{i=1}^N \left( \frac{\left( O_i-C_i \right) ^2}{ \sigma_i^2 } \right)
\end{equation}
\\
often called normalized chi-square, whose usage as goodness-of-fit estimator is very common in the literature. In all solutions presented here we give the values of both parameters.
\subsection{The M09 data set}
The initial observations of this series were used by Bonfils et al. (2005) to announce the discovery of GJ581 b. In a next paper, Udry et al (2007) reported the discovery of the planets GJ581 c and GJ581 d. In that paper, the periods of the 3 planets found, b, c and d were, respectively, $\approx 5.36$ d, $\approx 12.9$ d and {$\approx 83.4$ d}. The fourth planet, GJ581 e, was announced {in M09}, using the same 119 observations   considered here. In addition, the period of the planet d was recalculated as being $\approx 66$ days. Corrections of this kind are a common fact in the study of time series and are due to aliasing effects of the spacing of the data (all observations in one site being done near the same {sidereal time)}, when a sampling can represent more than one periodic function. The alias appears as a peak at a spurious frequency in the power spectrum. There is no mathematical tool to solve this kind of problem; the only possible solution is to have observations done at different sidereal times (or, equivalently, in sites well separated in longitude).
\\
Figure 1 shows the normalized DCDFT (Ferraz-Mello 1981) of the M09 data, after having eliminated from them the part of the signal coming from the planets b and c, by means of a {harmonic filtering. We see one peak} in the period $\approx 3.14$ d and also another one in $\approx 66$ d, corresponding respectively to {the planets e and d (M09).} The vertical dashed line represent the Nyquist frequency $\approx 0.5 d^{-1}$ where a mirror effect can be seen: the peaks in both sides of that line are nearly mirror images one from another. They are the result of the aliasing. This problem, for this system, was already detected by Fabrycky $\&$ Dawson {(2010)}. Figure 1 shows that the inspection of the power spectrum does not allow us to decide which alias corresponds to the actual period.
\\
The results of complete least-squares solutions (Table 1) show that this procedure is also not sufficient to decide among the two solutions corresponding to planet GJ581 d (but it allows us to decide between the periods $\approx 3.15$ and $\approx 1.47$ d for the planet e). {The dynamical analyses can be helpful in getting the correct period: numerical integrations of the conditions of the Table 1 are stable for a long period. However, we verify that the eccentricities of the planet d given by (1.1) and (1.2) are excited as soon as the integration begins. After that, the variation of this element stabilize in a value different of the initially proposed. This phenomenon does not occur with  (1.3) and (1.4) and it allows us conclude that the solutions with period $\approx 1.0123d$, in solutions (1.1) and (1.2), are not consistent from the dynamical point of view}.  
\\
We have also compared the model with circular orbits to one model with elliptic orbits. The improvement in the wrms when non-zero eccentricities are used is not statistically significant, given the increase in the number of degrees of freedom, as an F-test can show.
\begin{figure}
\begin{center}
 \includegraphics[width=10cm]{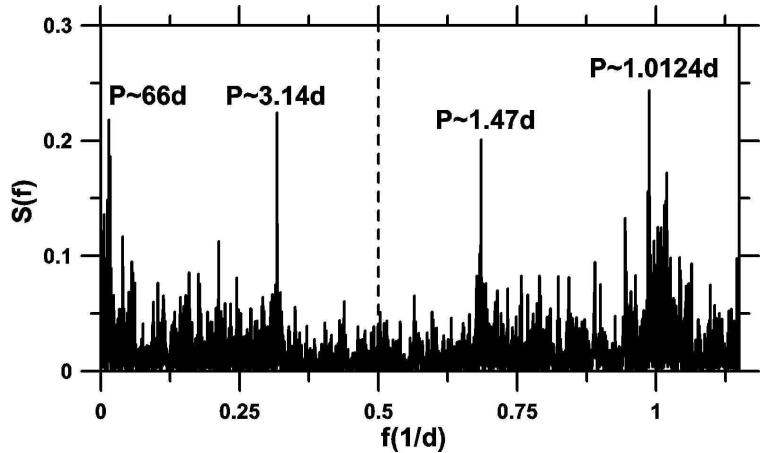} 
\caption{Normalized DCDFT of the M09 data after having eliminated from them the part of the signal coming from the planets b and c by means of a harmonic filtering, The vertical dashed line represent the Nyquist frequency $0.5 d^{-1}$.}
\label{fig:1}
\end{center}
\end{figure}
\begin{table}
\caption{Complete least-square solutions using the data in M09 considering, separately, the two possible periods for GJ581 d and two models with all planets in circular or eccentric orbits. In all solutions presented in this paper, the orbital longitude and argument of pericenter corresponds to the epoch JD-2451409.76222}
\label{tab:1}       
\begin{tabular}{llllllll}
\hline 
\bf{Solution 1.1} & $K(m/s)$ & $P(d)$ & $e$ & $\omega(\circ)$ & $\ell (\circ)$ & M ($M_{\otimes}$)   \\
\hline   
GJ581 b   & $12.5 \pm 0.6$ &  $5.3687 \pm 0.0005 $ & $0.001 \pm 0.07$ & $233.2 \pm 7$ & $253.4\pm 9$ & $15.7 \pm 0.3$ \\
GJ581 c   & $3.2 \pm 0.6$ &  $12.931 \pm 0.002 $ & $0.2 \pm 0.2$ & $267 \pm 50$ & $133.\pm 50$ & $5.3 \pm 0.2$ \\
GJ581 d   & $2.7 \pm 1.$ &  $1.0124 \pm 0.0001 $ & $0.37 \pm 0.3$ & $158 \pm 40$ & $285.\pm 40$ & $1.8 \pm 0.8$ \\
GJ581 e   & $1.67 \pm 0.7$ &  $3.149 \pm 0.002 $ & $0.18_{-0.1}^{0.5}$  & $133 \pm 35$ & $4.1\pm 30$ & $1.72_{-0.3}^{+0.7}$ \\
\hline
\multicolumn{7}{c}{$V_o=(-0.20 \pm 0.3)m/s$ \hspace{0.25cm} $Q=2.53$ \hspace{0.25cm} $wrms=1.54 m/s$  }\\ 
\hline
\hline
\bf{Solution 1.2} & $K(m/s)$ & $P(d)$ & $e$ & $\omega(\circ)$ & $\ell (\circ)$ & M ($M_{\otimes}$)   \\
\hline
GJ581 b   & $12.33 \pm 0.3$ &  $5.368 \pm 0.002 $ & $0.0 (fixed)$ & $0.0 (fixed)$ & $162.\pm 6$  & $15.5\pm 0.4$ \\
GJ581 c   & $2.8 \pm 1.7$ &  $12.93 \pm 0.03 $    & $0.0 (fixed)$ & $0.0 (fixed)$ & $20.\pm 25$  & $ 4.7\pm 2  $ \\
GJ581 d   & $2.5 \pm 1.5$ &  $1.0124 \pm 0.002$   & $0.0 (fixed)$ & $0.0 (fixed)$ & $311.\pm 30$ & $ 1.8\pm 0.2$ \\
GJ581 e   & $1.5 \pm 1.$ &   $3.15 \pm 0.001 $    & $0.0 (fixed)$ & $0.0 (fixed)$ & $204.\pm 20$ & $ 1.6\pm 0.4$ \\
\hline
\multicolumn{7}{c}{$V_o=(0.0 \pm 0.4)m/s$ \hspace{0.25cm} $Q=3.07$ \hspace{0.25cm} $wrms=1.77 m/s$ }\\
\hline
\hline
\bf{Solution 1.3} & $K(m/s)$ & $P(d)$ & $e$ & $\omega(\circ)$ & $\ell(\circ)$ & M ($M_{\otimes}$)   \\
\hline
GJ581 b   & $12.4 \pm 1.$ &  $5.3688 \pm 0.0007 $ & $0.02_{+0.3}^{-0.02} $ & $20 \pm 20 $ & $142.\pm 30$ & $ 15.5 \pm 0.5$ \\
GJ581 c   & $3.2 \pm 1.2$ &  $12.931 \pm 0.3    $ & $0.16_{+0.3}^{-0.1}  $ & $231.025   $ & $154.\pm 30$ & $ 5.3_{-1.7}^{+0.7} $ \\
GJ581 d   & $2.7 \pm 1.5$ &  $66.8 \pm 0.5      $ & $0.38 \pm 0.3        $ & $322 \pm 50$ & $107.\pm 50$ & $ 7.2 \pm 0.8$ \\
GJ581 e   & $1.9 \pm 1.6$ &  $3.157 \pm 0.08    $ & $0.11_{+0.7}^{-0.1}  $ & $145 \pm 50$ & $2.5 \pm 50$ & $ 2.0_{-1}^{1.7} $ \\
\hline
\multicolumn{7}{c}{$V_o=(-0.34 \pm 0.3)m/s$ \hspace{0.25cm} $Q=2.57$ \hspace{0.25cm}  $wrms=1.55 m/s$}\\
\hline
\hline
\bf{Solution 1.4} & $K(m/s)$ & $P(d)$ & $e$ & $\omega(\circ)$ & $\ell(\circ)$ & M ($M_{\otimes}$)   \\
\hline
GJ581 b   & $12.6 \pm 0.6$ &  $5.3688 \pm 0.0004 $ & $0.0 (fixed)$ & $0.0(fixed)$ & $162.\pm 6$ &  $15.8 \pm 0.2$ \\
GJ581 c   & $3.06 \pm 0.5$ &  $12.92 \pm 0.01 $    & $0.0 (fixed)$ & $0.0(fixed)$ & $123.\pm 10$ & $5.13 \pm 0.1$ \\
GJ581 d   & $2.21 \pm 0.5$ &  $66.74 \pm 0.4 $     & $0.0 (fixed)$ & $0.0(fixed)$ & $61.\pm 30$ &  $6.4 \pm  0.2$ \\
GJ581 e   & $1.75 \pm 0.6$ &  $3.15 \pm 0.015 $    & $0.0 (fixed)$ & $0.0(fixed)$ & $146.\pm 35$ & $1.8 \pm 0.2$ \\
\hline
\multicolumn{7}{c}{$V_o=(0.00 \pm 0.3)m/s$ \hspace{0.25cm}  $Q=2.91$ \hspace{0.25cm}  $wrms=1.72 m/s$}\\
\hline
\end{tabular}
\end{table}
\subsection{The V10 data set}

In V10, 109 measurements done at the Lick observatory over 11 years were published. With this data set alone we can infer only the presence of GJ581b, with the period of $\approx 5.4$ d, as can be seen in Table 2 and Figure 2. It is also not possible to obtain {a significantly better solution (F-test) using an elliptic orbit instead of a circular one}.
\\
The right panel of figure 2 shows the DCDFT of the residuals obtained after eliminating from the data the contribution of GJ581 b in an elliptic orbit. It shows that no important peak appears in the interval $0$ - $0.5 d^{-1}$ (It is enough to look at this interval because one possible peak with a higher frequency would produce an alias in this interval.) They are all below a minimum confidence level determined by taking  randomly shuffled sets of the data and seeing the height of the peaks that can be randomly formed; $15\%$ of the spectra have peaks higher than the level shown. Therefore, the V10 $'$ s data set alone does not allow other bodies to be detected in the system.

\begin{table}
\caption{Complete least-squares solutions using the data in V10 considering the planets GJ581 d in circular or eccentric motion, respectively.}
\label{tab:2}       
\begin{tabular}{llllllll}
\hline 
\bf{Solution 2.1} & $K(m/s)$ & $P(d)$ & $e$ & $\omega(\circ)$ & $\ell (\circ)$ & M ($M_{\otimes}$)   \\
\hline   
GJ581 b   & $12.25 \pm 0.6$ &  $5.369 \pm 0.001 $ & $0.0 (fixed) $ & $0.0 (fixed)$ & $286.\pm 17$ & $15.4 \pm 0.6$ \\
\hline
\multicolumn{7}{c}{$V_o=(1.11 \pm 0.5)m/s$ \hspace{0.25cm}  $Q=4.68$ \hspace{0.25cm}  $wrms=3.28 m/s$}\\
\hline
\bf{Solution 2.2} & $K(m/s)$ & $P(d)$ & $e$ & $\omega(\circ)$ & $\ell (\circ)$ & M ($M_{\otimes}$)   \\
\hline
GJ581 b   & $12.38 \pm 0.6$ &  $5.369 \pm 0.001 $ & $0.059 \pm 0.04$ & $43.5 \pm 10$ & $245.6\pm 10$ &  $15.5\pm 0.5$ \\
\hline
\multicolumn{7}{c}{$V_o==(1.01 \pm 0.5)m/s$ \hspace{0.25cm}  $Q=4.67$ \hspace{0.25cm}  $wrms=3.24 m/s$}\\
\hline
\end{tabular}
\end{table}

\begin{figure}
\begin{center}
  \includegraphics[width=16cm]{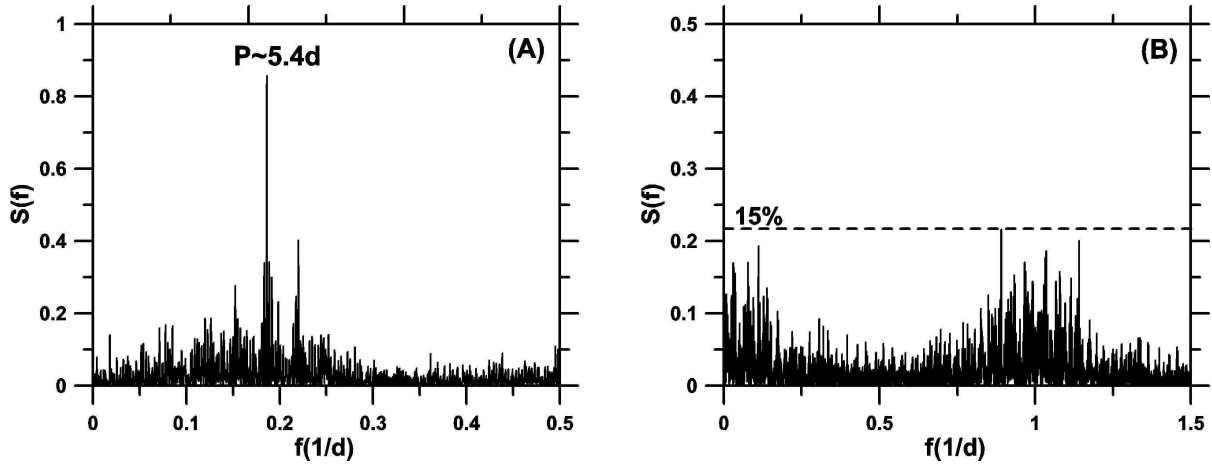}
\caption{(A) Normalized DCDFT spectra of the V10 data showing the peak corresponding to GJ581b, (B) DCDFT of the residuals. The dashed horizontal line shows the confidence level defined by shuffled data. }
\end{center}
\label{fig:2}
\end{figure}

\section{{The combined data set. GJ581 g}}

The discussions in the previous section have shown the limitation of the two existing sets of data. However, with the combined data set (V10+M09) it was possible to solve some of the difficulties discussed in previous section. It was possible to solve the alias problem thanks to the $\approx 3$h difference in longitude of the two observatories. The joint V10+M09 data did also allowed Vogt {et al. (2010)} to claim the discovery of two new bodies: GJ581 f and GJ581 g, with periods $\approx 433$ d and $\approx 36$ d respectively. The claimed planet GJ581 g lays
in the habitable zone, which, for the star GJ581, is in the range  $[0.11-0.21]$ AU ({Von Braun et al.(2011)}).

\subsection{Fourier Analysis}

Again, by harmonic filtering, we remove from the combined data the contribution of the planets b and c and Fourier analyze the residuals. The result is shown in Figure 3. The highest peak is now the one in the period {$\approx 67$ d}, but we also see the peak corresponding to the period $\approx 3.14$ d {(planet e)} and their aliases.

\begin{figure}
\begin{center}
  \includegraphics[width=8.cm]{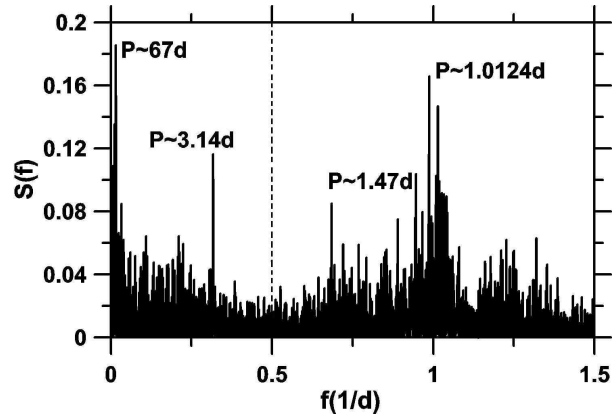}
\caption{ Normalized DCDFT of the joint M09+V10 data after elimination of the contributions due to planets b and c. The vertical dashed line represent the Nyquist frequency $\approx 0.5d^{-1}$.}
\end{center}
\label{fig:3}
\end{figure}

As before, it is not yet possible to choose from the spectrum which one is the true period of planet d. A detailed analysis of all possible solutions is still necessary. In Table 3, we give the complete least-squares orbital solutions for the combined data set with 4 planets considering the two possible periods for planet d. Let us stress the fact that in these solutions the periods are not fixed a priori but result from the least-squares best-fit procedure used. We just bracket the interval where the solution is searched. In one case we searched for a solution in the interval $[40-80]$ d while in the other we searched it in the interval $[0.5-10]$ d. {In Table 3 we} show that the difference in the resulting wrms, in both cases, is significant allowing us to decide in favor of the longest of the two periods. One may note, however, that the differences between the wrms of the models with circular and elliptical orbits still does not make possible to decide in favor of one model or another.
\\
The spectra of the residuals of the solutions given in {Table 3} are shown in {Figure 4}, where it is possible to devise the existence of GJ581 g, in the residuals of the circular model. It is also shown that such a period does not appear in the power spectrum of the residuals of the eccentric model.
\begin{table}
\caption{Least-squares solutions considering four bodies and the joint data sets M09 and V10.}
\label{tab:3} 
\begin{tabular}{llllllll}
\hline 
\bf{Solution 3.1} & $K(m/s)$ & $P(d)$ & $e$ & $\omega(\circ)$ & $\ell (\circ)$ & M ($M_{\otimes}$)   \\
\hline 
GJ581 b   & $12.53 \pm 0.9$ &  $5.369  \pm 0.002 $ & $0.0 (fixed)$ & $0.0 (fixed)$ & $287. \pm 10$ & $15.7 \pm 0.3$ \\
GJ581 c   & $3.01  \pm 0.7$ &  $12.902 \pm 0.008 $ & $0.0 (fixed)$ & $0.0 (fixed)$ & $35.  \pm 27$ & $3.8 \pm 0.3$ \\
GJ581 d   & $1.9   \pm 1. $ &  $66.81  \pm 0.4   $ & $0.0 (fixed)$ & $0.0 (fixed)$ & $36.  \pm 12$ & $5.5 \pm 2$ \\
GJ581 e   & $1.47  \pm 0.8$ &  $3.149  \pm 0.8   $ & $0.0 (fixed)$ & $0.0 (fixed)$ & $300. \pm 40$ & $1.5 \pm 0.5$ \\
\hline
\multicolumn{7}{c}{$V_{(M09)}=(-0.42 \pm 0.5) m/s$ \hspace{0.25cm} $V_{(V10)}=(1.1 \pm 1.5)  m/s $ \hspace{0.25cm} $Q=3.13$ \hspace{0.25cm} $wrms=2.11m/s$ }\\
\hline
\hline
\bf{Solution 3.2} & $K(m/s)$ & $P(d)$ & $e$ & $\omega(\circ)$ & $\ell (\circ)$ & M ($M_{\otimes}$)   \\
\hline
GJ581 b   & $12.5 \pm 0.4$ &  $5.369 \pm 0.005 $ & $0.002_{-0.0001}^{+0.01} $ & $44  \pm 30$ & $243 \pm 30$ &  $15.7_{-3}^{+0.3}$ \\
GJ581 c   & $2.8  \pm 0.8$ &  $12.92 \pm 0.05  $ & $0.01_{-0.01}^{+0.3}     $ & $321 \pm 50$ & $96  \pm 30$ &  $4.7_{-1.3}^{+0.3}$ \\
GJ581 d   & $2.7  \pm 1.2$ &  $66.9  \pm 0.6   $ & $0.54_{-0.5}^{+0.2}      $ & $324 \pm 30$ & $117 \pm 30$ &  $6.56_{-4}^{+0.4}$ \\
GJ581 e   & $1.6  \pm 1. $ &  $3.15  \pm 0.2   $ & $0.21_{-0.2}^{0.6}       $ & $132 \pm 30$ & $182 \pm 30$ &  $1.64_{-0.3}^{0.6}$ \\
\hline
\multicolumn{7}{c}{$V_{(M09)}=(-0.34 \pm 0.9) m/s$ \hspace{0.25cm} $V_{(V10)}=(1. \pm 1.)  m/s $ \hspace{0.25cm} $Q=3.01$ \hspace{0.25cm} $wrms=2.03m/s$ }\\
\hline
\hline
\bf{Solution 3.3} & $K(m/s)$ & $P(d)$ & $e$ & $\omega(\circ)$ & $\ell (\circ)$ & M ($M_{\otimes}$)   \\
\hline
GJ581 b   & $12.4 \pm 0.5$ &  $5.368  \pm 0.004  $ & $0.0 (fixed)$ & $0.0 (fixed)$ & $284.\pm 8$  & $15.5 \pm 0.5 $ \\
GJ581 c   & $2.73 \pm 0.8$ &  $12.93  \pm 0.01   $ & $0.0 (fixed)$ & $0.0 (fixed)$ & $51. \pm 40$ & $4.6 \pm 0.2  $ \\
GJ581 d   & $1.64 \pm 0.6$ &  $1.012  \pm 0.001  $ & $0.0 (fixed)$ & $0.0 (fixed)$ & $72. \pm 18$ & $1.2 \pm 0.2  $ \\
GJ581 e   & $1.3  \pm 1. $ &  $3.15   \pm 0.02   $ & $0.0 (fixed)$ & $0.0 (fixed)$ & $42. \pm 30$ & $1.4 \pm 0.6  $ \\
\hline
\multicolumn{7}{c}{$V_{(M09)}=(-0.5 \pm 1.) m/s$ \hspace{0.25cm} $V_{(V10)}=(0.8 \pm 2.)  m/s $ \hspace{0.25cm} $Q=3.17$ \hspace{0.25cm} $wrms=2.18m/s$ }\\
\hline
\hline
\bf{Solution 3.4} & $K(m/s)$ & $P(d)$ & $e$ & $\omega(\circ)$ & $\ell (\circ)$ & M ($M_{\otimes}$)   \\
\hline
GJ581 b   & $12.4 \pm 1.$ &  $5.368  \pm 1.     $ & $0.004_{-0.001}^{+0.2}      $ & $202.0 \pm 30  $ & $82  \pm 40$ &  $15.5\pm 1$ \\
GJ581 c   & $2.9  \pm 1.$ &  $12.92  \pm 0.04   $ & $0.2_{-0.15}^{+0.5}         $ & $179.0 \pm 30  $ & $206 \pm 30$ &  $4.8_{-1.2}^{+0.7}$ \\
GJ581 d   & $1.8  \pm 1.$ &  $1.012  \pm 0.003  $ & $0.2_{-0.2}^{+0.6}          $ & $3.0   \pm 27  $ & $69  \pm 30$ &  $1.3_{-0.7}^{+0.3}$ \\
GJ581 e   & $1.4  \pm 1.$ &  $3.15   \pm 0.5    $ & $0.14_{-0.1}^{+0.7}         $ & $104.0 \pm 100 $ & $104 \pm 100$&  $1.45\pm 0.5$ \\
\hline
\multicolumn{7}{c}{$V_{(M09)}=(-0.4 \pm 0.9) m/s$ \hspace{0.25cm} $V_{(V10)}=(0.8 \pm 1.)  m/s $ \hspace{0.25cm} $Q=3.66$ \hspace{0.25cm} $wrms=2.23m/s$ }\\
\hline
\end{tabular}
\end{table}
\begin{figure}
\begin{center} 
  \includegraphics[width=16cm]{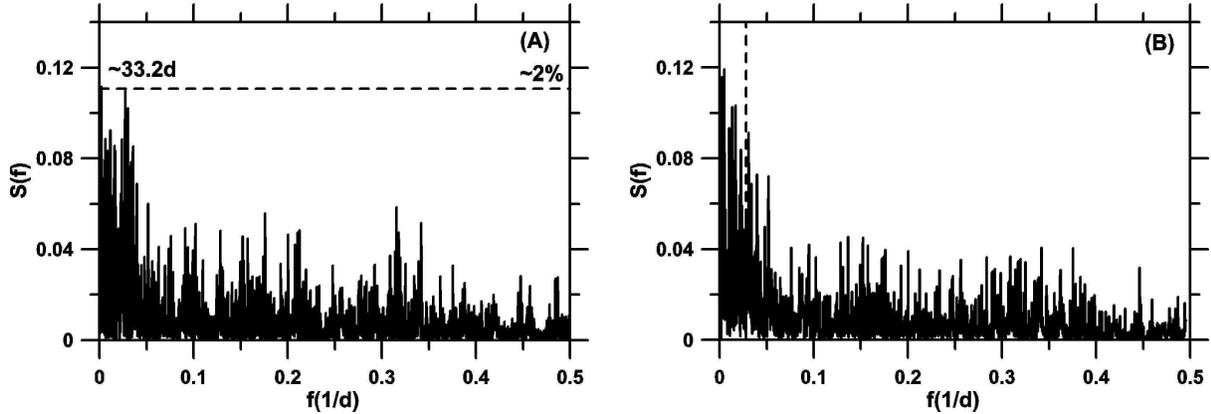}
\caption{ (A) Normalized DCDFT spectrum of the residuals of the circular solution (3.1), showing the peak corresponding to GJ581 g at ~33.2 d. (B) Normalized DCDFT spectrum of the residuals of the eccentric solution (3.2). In this case, the peak corresponding to GJ581 g is not seen (its position is represented with a vertical dashed line). The horizontal dashed lines represent the minimum level of confidence of the peaks.}
\end{center}
\label{fig:4}
\end{figure}
\begin{table}
\caption{Best-fit solution of the model with 5 planets in circular orbits. The fifth planet included in the analysis is GJ 581 g. The longitude of planet g is not constrained, so we can get good fits in all longitudes, the given value corresponds to the best fit of the period.}
\label{tab:4} 
\begin{tabular}{llllllll}
\hline 
\bf{Solution 4.1} & $K(m/s)$ & $P(d)$ & $e$ & $\omega(\circ)$ & $ \ell (\circ)$ & M ($M_{\otimes}$)   \\
\hline 
GJ581 b   & $12.96 \pm 0.5$ &  $5.3687 \pm 0.0002 $ & $0.0 (fixed) $ & $0.0 (fixed) $ & $288.\pm 9 $ & $16.3 \pm 0.2$ \\
GJ581 c   & $2.8   \pm 1. $ &  $12.92  \pm 0.01   $ & $0.0 (fixed) $ & $0.0 (fixed) $ & $66. \pm 25$ & $4.7 \pm 1$ \\
GJ581 d   & $2.1   \pm 1. $ &  $66.9   \pm 0.5    $ & $0.0 (fixed) $ & $0.0 (fixed) $ & $78. \pm 20$ & $6.1 \pm 3$ \\
GJ581 e   & $1.6   \pm 0.8$ &  $3.15   \pm 0.01   $ & $0.0 (fixed) $ & $0.0 (fixed) $ & $53. \pm 30$ & $1.7 \pm 0.3$ \\
GJ581 g   & $1.0   \pm 0.6$ &  $35.5   \pm 4.     $ & $0.0 (fixed) $ & $0.0 (fixed) $ & $67.       $ & $2.3 \pm 0.6$ \\
\hline
\multicolumn{7}{c}{$V_{(M09)}=(-0.3 \pm 0.9) m/s$ \hspace{0.25cm} $V_{(V10)}=(0.9 \pm 1.)  m/s $ \hspace{0.25cm} $Q=3.06$ \hspace{0.25cm} $wrms=2.15m/s$ }\\
\hline
\end{tabular}
\end{table}

\subsection{The Biased Monte Carlo (BMC)}

The BMC method consists in applying incomplete least square fits to intial guesses chosen at random in a given domain of the space of parameters. It was first used in Ferraz-Mello et al.{(2005a)} to map the wrms values in the neighborhood of the minimum and serves a double purpose: on one hand, the map thus obtained allow us to define the confidence interval of these results; on the other hand, it allows us to check if the minimum found by the optimization procedure is not just one among others. Each random guess is propagated running the optimization algorithms for a pre-determined number of iterations. The results farther from the minimum than a given amount are discarded. This is the case for many of the runs. Given the large number of unknowns, one pure random process is not feasible as the probability of getting one point near the minimum by pure chance is almost zero. The parameters of the incomplete optimization are set to be such that the procedure gives a reasonable amount of sampled points near the least squares solution and allows us to map the wrms on a significantly wide domain of the phase space. The solutions thus obtained are called 'good-fit solutions'. They are not mathematically as good as the best-fit, but they are statistically equivalent to it, since they belong to the confidence domain of the best fit. We use the BMC technique here in order to verify if each one of the orbital elements are well or poorly determined. It makes possible a better comparison between models and also allow us to seek for secondary minima, thus improving the results of the previous section. Therefore, we take the V010+M09 data set and look for good fits of the model with four planets in circular and eccentric orbits. Afterwards we compare this results with those obtained after the inclusion of the planet g.
\\
We have thus applied the BMC technique to four different models: four planets with all orbits circular or with two of them eccentric (as described below) and including or not the fifth planet in circular orbit. {These results} are shown in {Figure 5}. In panel (A), black dots indicate the good fits obtained when we consider four planets, two of them, GJ581 d and e, with periods 66 d and 3.15 d, in eccentric orbits and {the other two, GJ581 b and c, in circular orbits. We remind that in the solution 3.2, the orbits of these two planets are almost circular. The innermost planet (GJ581 e) could also be assumed in circular orbit (see Papaloizou, 2011) but was kept as elliptic because of the nonzero value found for its eccentricity in solution 3.2}. The horizontal black dashed line represents the best solution found, with $wrms=2.08 m/s$. In the same panel, red dots indicate the results when a fifth planet, {in circular orbit}, is added to the model, in which case the best-fit solution is {obtained at} $wrms=2.06 m/s$. So, from (A), it is not possible to see any statistical difference between the two considered models and the inclusion of the planet GJ581 g in this case is unjustified, as it does not bring any improvement to the wrms of the best-fit solutions. The best-fit solutions with 5 circular planets and with 4 planets (2 in circular and 2 in ellipitic orbits) are given in Table 5.
\\
In panel (B), we have a similar situation but considering all planets in circular orbits. Now the wrms of the best-fit solution in the model with 4 planets is $2.17m/s$ and that of the model with 5 planets is $2.03 m/s$. In this case, the inclusion of the fifth planet (GJ581 g) means a real improvement and would support a conclusion {positive concerning the existence of the planet GJ581 g. The best-fit solutions with 5 circular planets and with 4 planets (2 in circular and 2 in ellipitic orbits) are given in Table 5.}\\
\\
Figure 6 shows BMC-fits of some orbital elements of GJ581 g using the model with circular orbits. It {allows} us to estimate the quality of the solution given in Table 5.1. The best value of the amplitude is {$K_g=1.2 m/s$}. {The period $P_g$}, however, shows two minima, $\approx 33$ d and $\approx 36$ d, which appear represented with {the labels $P'$ and $P$}. This difference is due to an aliasing indetermination due to observations concentrated near the oppositions. We indeed have {$1/33d \approx (1/36d)+(1/365d)$. At} last, we add that the orbital phase at the epoch, $\ell_g$ , is not constrained as equally good fits are found at all longitudes. This is a consequence of the large interval of the confidence for the period of GJ581g. Variations in the period inside this interval are enough to give values of the longitude at epoch over all possible values.

\begin{figure}
\begin{center}
  \includegraphics[width=16cm]{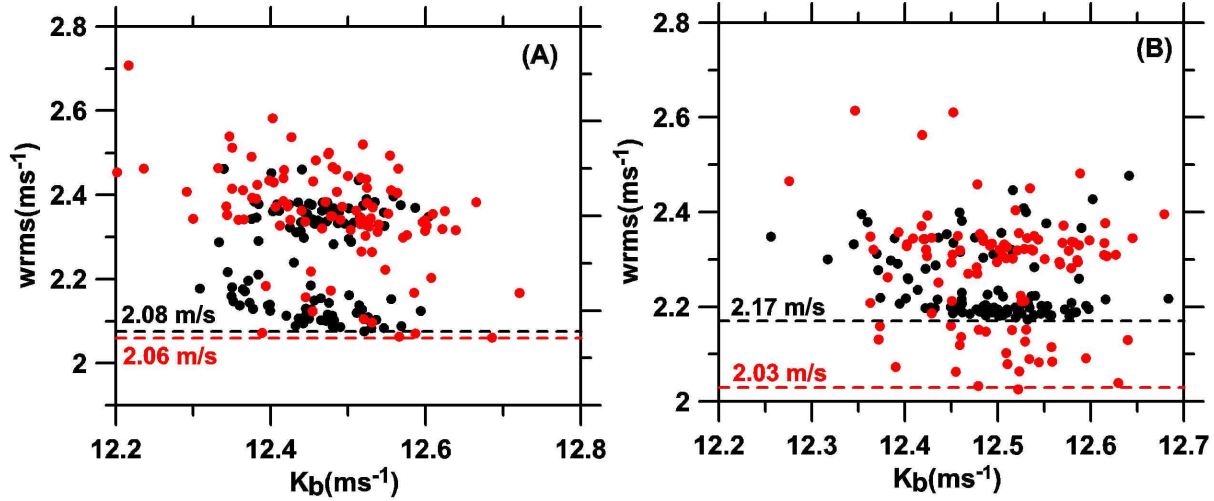}
\caption{Comparision of BMC fits of differents models: (A) the planets b, c and g are in circular orbits and d and e in eccentric orbits and in (B) all planets are in circular orbits. Red and black points are the results obtained when the planet g is respectively included or not included in the model. The horizontal dashed lines show the minimal values of the wrms (best-fit) reached in each case.}
\label{fig:5}
\end{center}
\end{figure}

\begin{table}
\caption{{Best fit solution with 5 planets in circular orbits (5.1) and with 4 planets, two of which in eccentric orbits (5.2)}.
In these solutions, the longitudes of the planets e and g are not constrained. The given value corresponds to the nominal period.}
\label{tab:5} 
\begin{tabular}{llllllll}
\hline 
\bf{Solution 5.1} & $K(m/s)$ & $P(d)$ & $e$ & $\omega(\circ)$ & $\ell (\circ)$ & M ($M_{\otimes}$)   \\
\hline 
GJ581 b   & $12.5 \pm 0.2        $ &  $5.369      \pm 0.001  $ & $0.0 (fixed)$ & $0.0 (fixed)$ & $288. \pm 30 $  & $15.7 \pm 0.7$ \\
GJ581 c   & $2.92 \pm 0.6        $ &  $12.9       \pm 0.1    $ & $0.0 (fixed)$ & $0.0 (fixed)$ & $22.5 \pm 27 $  & $4.9 \pm 0.1$ \\
GJ581 d   & $2.00 \pm 0.6        $ &  $66.65_{-0.6}^{1}      $ & $0.0 (fixed)$ & $0.0 (fixed)$ & $3.8  \pm 30 $  & $5.6 \pm 1$ \\
GJ581 e   & $1.5_{-1.0}^{+0.3}   $ &  $3.15_{-2}^{0.8}       $ & $0.0 (fixed)$ & $0.0 (fixed)$ & $300.         $ & $1.7_{-1}^{+0.4}$ \\
GJ581 g   & $1.19_{-1.0}^{+0.2}  $ &  $36.5_{-9}^{3}         $ & $0.0 (fixed)$ & $0.0 (fixed)$ & $248.         $ & $2.8_{-2}^0.2$ \\
\hline
\multicolumn{7}{c}{$V_{(M09)}=(-0.6 \pm 0.6) m/s$ \hspace{0.25cm} $V_{(V10)}=(1.1 \pm 0.4)  m/s $ \hspace{0.25cm} $Q=3.23$ \hspace{0.25cm} $wrms=2.03m/s$ }\\
\hline
\hline
\bf{Solution 5.2} & $K(m/s)$ & $P(d)$ & $e$ & $\omega(\circ)$ & $\ell (\circ)$ & M ($M_{\otimes}$)   \\
\hline
GJ581 b   & $12.5 \pm 0.2$    &  $5.369  \pm 0.001 $ & $0.0 (fixed)         $ & $0.0 (fixed) $ & $287.\pm 10$ &  $15.7\pm 0.7$ \\
GJ581 c   & $3.   \pm 0.7$    &  $12.92  \pm 0.02  $ & $0.0 (fixed)         $ & $0.0 (fixed  $ & $9.6 \pm 27$ &  $5.0 \pm 0.9$ \\
GJ581 d   & $2.5  \pm 0.5$    &  $66.8   \pm 0.3   $ & $0.4_{-0.2}^{+0.12}  $ & $331 \pm 50  $ & $69. \pm 50$ &  $6.6_{-1}^{+0.4}$ \\
GJ581 e   & $1.6_{-1}^{+0.2}$ &  $3.15_{-2}^{+0.5} $ & $0.01_{-0.01}^{+0.4} $ & $235         $ & $78.       $ &  $1.7_{-1.6}^{+0.7}$ \\
\hline
\multicolumn{7}{c}{$V_{(M09)}=(-0.3 \pm 0.4) m/s$ \hspace{0.25cm} $V_{(V10)}=(0.9 \pm 0.5)  m/s $ \hspace{0.25cm} $Q=2.9$ \hspace{0.25cm} $wrms=2.08m/s$ }\\
\hline
\end{tabular}
\end{table}
Figure 7 is the analogous to {Figure 5} obtained when only the data in the set M09 are considered. We see that, in this case, the improvement of the results when the fifth planet is included is not significant. We have found solutions with periods in the range [25-36] d, but in all cases it was not possible to get a robust inference about this additional body. The plot for amplitude and phase are similar to those shown in {Figure 6}. The smaller wrms of the solutions of Table 1 results from the homogeneity of the set M09. It is important to stress that the wrms is an evaluation of the statistical errors and {does not include} possible systematic differences between the two data sets.

\begin{figure}
\begin{center}
  \includegraphics[width=16cm]{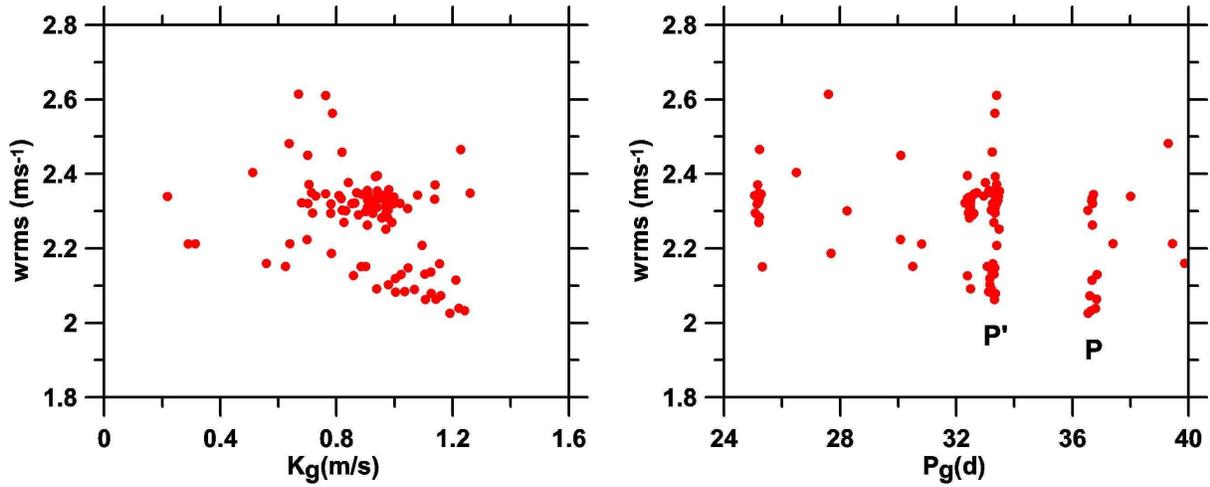}
\caption{BMC-fits of the orbital parameters of GJ581 g in the model of the 5 planets in circular orbits.}
\label{fig:6}
\end{center}
\end{figure}

\begin{figure}
\begin{center}
  \includegraphics[width=16cm]{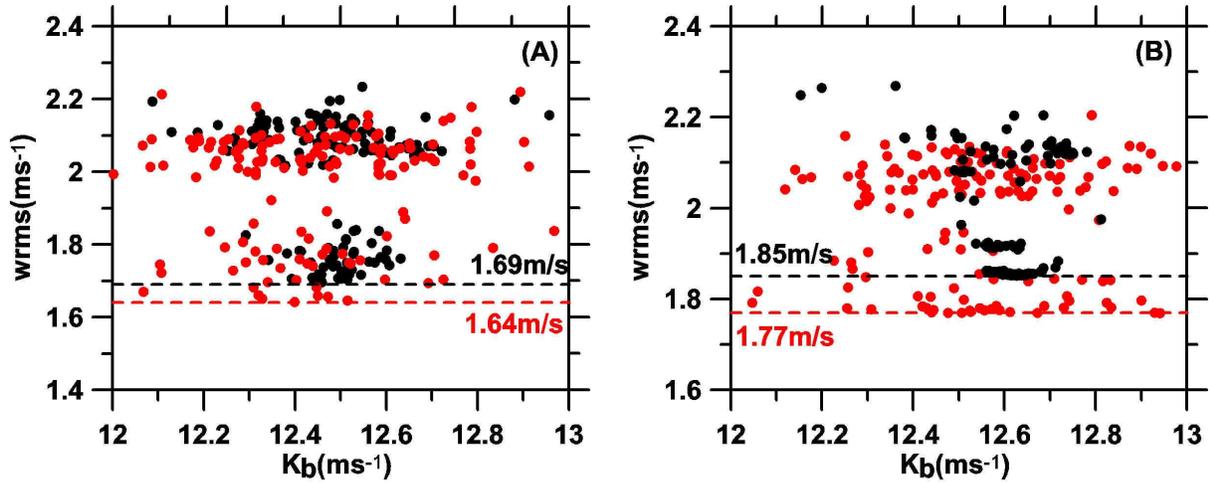}
\caption{Same as {Figure 5} when considering only the M09 data set.}
\label{fig:7}
\end{center}
\end{figure}
{Solutions initially with 5 planets in circular orbits, or with only 4 of them, but in elliptic orbits with GJ581 d initially in a very eccentric orbit $(e=0.43)$, were simulated using precise numerical integrations for times up to 300 Myr. The solutions found did not show any appreciable variation in the considered time span. They also appeared to be very robust with respect to small variations in the initial conditions.}

\subsection{The dynamics in the Habitable Zone}

In this section, we give some dynamical constraints concerning the existence of one body between the planets GJ581 c and d, using for the 4 planets the values of the solution 5.1. {The mass and the initial longitude of the fifth planet were taken from solution 5.1. Its semi-axis and eccentricity were taken on a grid in the interval:} $a_g=[0.1-0.18]$, $\Delta_a=0.002$ AU  and $e_g=[0.0-0.5]$, $\Delta_e=0.0025$. Each point of the grid was integrated for $10^5 yrs$ and we calculated {Michtchenko$'$s spectral numbers associated with the evolutions of its eccentricity and semi-major axis (see Ferraz-Mello et al., 2005). Solutions are considered unstable (chaotic) when the spectral number is high ($\log N \geq 1.8$) or when the test planet crosses the orbit of one of the neighboring planets after leaving the Habitable Zone.}  These situations are represented in the map  by the red painted areas. The blue and purple regions represent the domains of regular motion.{The white/black mark shows the position of planet GJ581 g in solution 5.1 (the eccentricity was taken as 0.001 as it is expected to be due to the forcing by the other planets).}

\begin{figure}
\begin{center}
  \includegraphics[width=16cm]{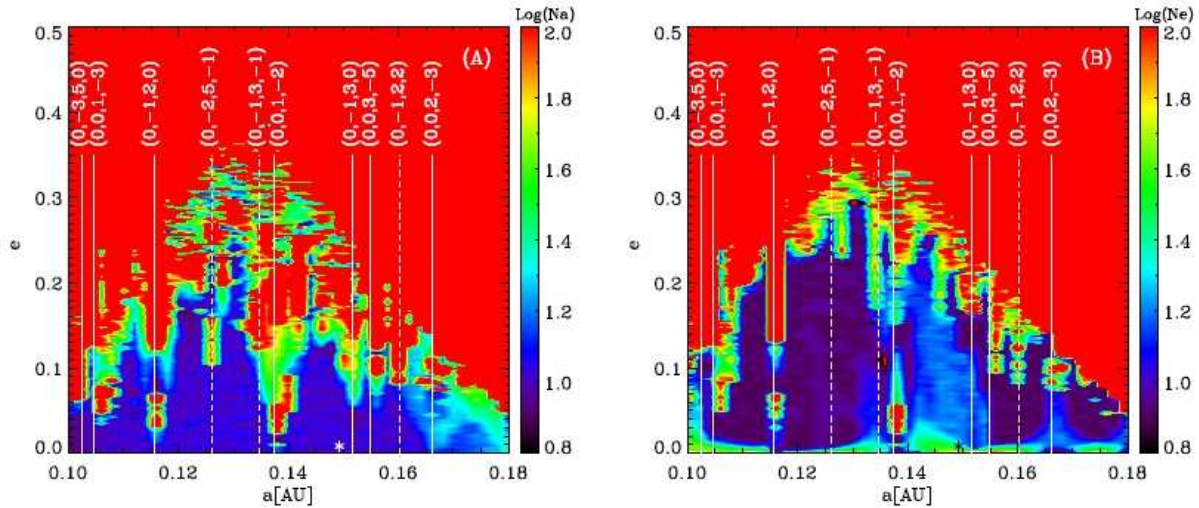}
\caption{Dynamical map in the region of the planet g, calculated with the initial conditions and masses of solution 5.1 . In (A) and (B) we have the log of spectral number of semimajor-axis and eccentricity, respectively.}
\label{fig:8}
\end{center}
\end{figure}

{Figure (8) shows that several mean-motion resonances (MMR) act as source of instability in the Habitable Zone. The location of some selected resonances coinciding with the chaotic regions appearing on the map is shown by vertical lines. They are two- and three-body resonances involving the fifith planet and one of the neighboring planets, GJ581 b, c and d. They may be generally written as}
\begin{equation}
k_b n_b + k_c n_c + k_t n_t + k_d n_d \simeq 0 . 
\end{equation}
\\ 
{We selected those corresponding to the main instabilities appearing in the maps. Their coefficients are indicated by the n-uples $(k_b,k_c,k_t,k_d)$. Among them, we point out the resonances $(0,0,1,-2)$ (i.e. $n_t-2n_d \simeq 0$) at $a \simeq 0.138$ AU, and $(0,1,-2,0)$ (i.e. $n_c-2n_t \simeq 0$) at $a \simeq 0.116$ AU, as the most important, affecting the solutions at all eccentricities. It is also worth noting that no low-order resonances were found near some importante features of the maps, which appear to be associated with three-body resonances. Some of them exhibit a highly chaotic dynamics (at moderate-to-low-eccentricities) which may be explained by the formation of resonances multiplets (see Cachucho et al. 2010). One such multiplet was found around the resonance $(0,-1,3-1)$ (i.e. $-n_c+3n_t-n_d \simeq 0$ at $a \simeq 0.135$ AU) shown in Figure 8. The formation of such overlapping multiplets is responsible for diffusion across the resonances and the more intense chaotic behavior shown by the semi-majo axis (Figure 8A).} \\
{At last, we have to stress the fact that a similar study using the eccentric solutions 5.2 lead only to unstable solutions. This shows that the eccentricity of GJ581 d plays a crucial role in the stability of a planet placed in the Habitable Zone. The high values for its eccentricity found here and in other determinations ($e \simeq 0.4-0.5$) make dinamically impossible the existence of any other body in the HZ. In that case, the perihelic distance of GJ581 d lies in the middle of the HZ and is responsible for the rapid ejection of all test bodies placed there.}

\section{The planet f}

Figure 9 shows the normalized DCDFT of the residuals of the solution (5.1) given in Table 5. One {may see a peak} corresponding to the period $\approx 455$ d. However, its confidence level, estimated by comparing shuffled data spectra is $\simeq 4\%$.
%\textbf{In the literature it is common to consider} as reliable periodicity that one with peaks in the confidence level $\leq 3\%$.
So, the planet f is in the limit of detection. For the sake of completeness, we give in {Table 6} the orbital elements obtained with 6 circular orbits.
\begin{figure}
\begin{center}
  \includegraphics[width=10cm]{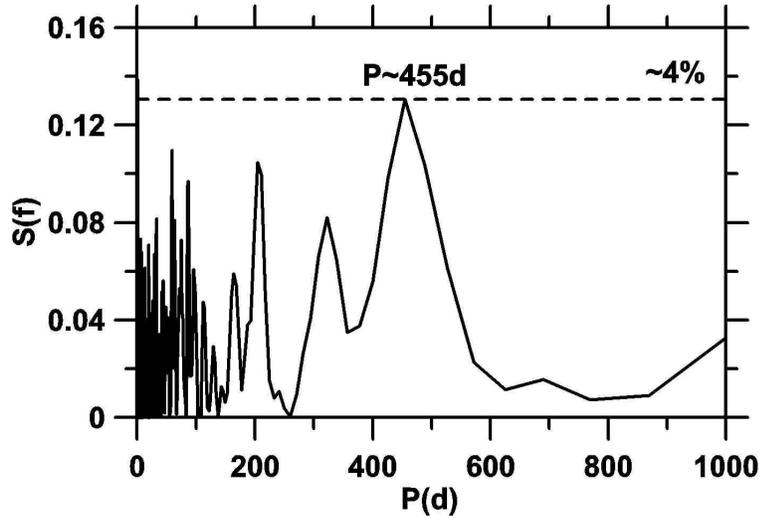}
\caption{Normalized DCDFT of the residuals {of the solution 5.1}. The dashed horizontal line is the minimum level of confidence of the peaks.}
\label{fig:9}
\end{center}
\end{figure}
\begin{table}
\caption{Best-fit circular solution with 6 planets. The longitudes of planets e,g and f are not constrained.}
\label{tab:6} 
\begin{tabular}{llllllll}
\hline 
\bf{Solution 6.1} & $K(m/s)$ & $P(d)$ & $e$ & $\omega(\circ)$ & $\ell (\circ)$ & M ($M_{\times}$)   \\
\hline 
GJ581 b   & $12.5 \pm 0.6        $ &  $5.3687 \pm 0.001  $ & $0.0 (fixed)$ & $0.0 (fixed)$ & $288 \pm 5 $ & $15.7 \pm 0.3  $ \\
GJ581 c   & $2.7  \pm 1.         $ &  $12.9   \pm 0.4    $ & $0.0 (fixed)$ & $0.0 (fixed)$ & $71  \pm 30$ & $4.5 \pm 1.5   $ \\
GJ581 d   & $2.4  \pm 0.7        $ &  $66.8   \pm 0.3    $ & $0.0 (fixed)$ & $0.0 (fixed)$ & $64  \pm 27$ & $6.9 \pm 1     $ \\
GJ581 e   & $1.6  \pm 1.         $ &  $3.15   \pm 0.01   $ & $0.0 (fixed)$ & $0.0 (fixed)$ & $25        $ & $1.7 \pm 0.3   $ \\
GJ581 f   & $1.1 \pm  0.8        $ &  $33.3   \pm 4.     $ & $0.0 (fixed)$ & $0.0 (fixed)$ & $44        $ & $2.6 \pm 1.5   $ \\
GJ581 g   & $0.9_{-0.8}^{+0.5}   $ &  $425.0_{-25}^{+65} $ & $0.0 (fixed)$ & $0.0 (fixed)$ & $70        $ & $4.6_{-4}^{2.4}$ \\
\hline
\multicolumn{7}{c}{$V_{(M09)}=(-0.5 \pm 1.) m/s$ \hspace{0.25cm} $V_{(V10)}=(0.7 \pm 0.7)  m/s $ \hspace{0.25cm} $Q=2.52$ \hspace{0.25cm} $wrms=1.95m/s$ }\\
\hline

\end{tabular}
\end{table}
\section{Conclusion}

From a purely statistical point of view, it is not incorrect to state the existence of GJ581 g. However, this can only be done by assuming that all planets are in circular orbits. The explanation of this fact is a well-known phenomenon, described in detail in {Anglada-Escud\'e et al. (2009, 2010)} and Giuppone et al. (2009), happens when the periods are close to a commensurability relation (the period of GJ581 d is almost twice the period of GJ581 g). In this circumstance, the eccentricity of the external body will be overestimated and can hide totally the signal of the internal body. In this case, the analysis of the radial velocity measurements does not allow us to distinguish the signal of the putative fifth planet (GJ581 g) from an overestimated eccentricity of the orbit of GJ581 d. Additional tests were done, but we have not been able to verify the existence of the planet g when we consider the d with a fixed and low ($\approx 0.1$) but not zero eccentricity.\\
Of course, additional observations can help solving the dilemma about planet g. {Forveille et al. 2011}, using new observations, were not able to confirm the existence of the planets g and f. But, this statement could not be verified because the new data were not yet made available in public domain. In adition, we have that the analysis done by them {did not include} the V10 data set.\\
The signal with the frequency of the alleged sixth planet - GJ581 f - was found, but it is in the threshold of confidence level.{We cannot discard the hypotesis that the period of the sixth planet is produced by some complicated beating with the observation window of 1yr, once both are comparables.} 
\newpage
\section{Summary}

Our best-fit with 4 planets in eccentric orbits (Solution 3.2 in the text) is:

\begin{table}[h]
\begin{tabular}{llllllll}
\hline
\bf{} & $K(m/s)$ & $P(d)$ & $e$ & $\omega(\circ)$ & $\ell (\circ)$ & M ($M_{\otimes}$)   \\
\hline
GJ581 e   & $1.6  \pm 1. $ &  $3.15  \pm 0.2   $ & $0.21_{-0.2}^{0.6}       $ & $132 \pm 30$ & $182 \pm 30$ &  $1.64_{-0.3}^{0.6}$ \\
GJ581 b   & $12.5 \pm 0.4$ &  $5.369 \pm 0.005 $ & $0.002_{-0.0001}^{+0.01} $ & $44  \pm 30$ & $243 \pm 30$ &  $15.7_{-3}^{+0.3}$ \\
GJ581 c   & $2.8  \pm 0.8$ &  $12.92 \pm 0.05  $ & $0.01_{-0.01}^{+0.3}     $ & $321 \pm 50$ & $96  \pm 30$ &  $4.7_{-1.3}^{+0.3}$ \\
GJ581 d   & $2.7  \pm 1.2$ &  $66.9  \pm 0.6   $ & $0.54_{-0.5}^{+0.2}      $ & $324 \pm 30$ & $117 \pm 30$ &  $6.56_{-4}^{+0.4}$ \\
\hline
\multicolumn{7}{c}{$V_{(M09)}=(-0.34 \pm 0.9) m/s$ \hspace{0.25cm} $V_{(V10)}=(1. \pm 1.)  m/s $ \hspace{0.25cm} $Q=3.01$ \hspace{0.25cm} $wrms=2.03m/s$ }\\
\hline
\end{tabular}
\end{table}

The best-fit with 5 planets in circular orbits (Solution 4.1 in the text) is: 

\begin{table}[h]
\begin{tabular}{llllllll}
\hline 
 & $K(m/s)$ & $P(d)$ & $e$ & $\omega(\circ)$ & $ \ell (\circ)$ & M ($M_{\otimes}$)   \\
\hline 
GJ581 e   & $1.6   \pm 0.8$ &  $3.15   \pm 0.01   $ & $0.0 (fixed) $ & $0.0 (fixed) $ & $53. \pm 30$ & $1.7 \pm 0.3$ \\
GJ581 b   & $12.96 \pm 0.5$ &  $5.3687 \pm 0.0002 $ & $0.0 (fixed) $ & $0.0 (fixed) $ & $288.\pm 9 $ & $16.3 \pm 0.2$ \\
GJ581 c   & $2.8   \pm 1. $ &  $12.92  \pm 0.01   $ & $0.0 (fixed) $ & $0.0 (fixed) $ & $66. \pm 25$ & $4.7 \pm 1$ \\
GJ581 g   & $1.0   \pm 0.6$ &  $35.5   \pm 4.     $ & $0.0 (fixed) $ & $0.0 (fixed) $ & $67.       $ & $2.3 \pm 0.6$ \\
GJ581 d   & $2.1   \pm 1. $ &  $66.9   \pm 0.5    $ & $0.0 (fixed) $ & $0.0 (fixed) $ & $78. \pm 20$ & $6.1 \pm 3$ \\
\hline
\multicolumn{7}{c}{$V_{(M09)}=(-0.3 \pm 0.9) m/s$ \hspace{0.25cm} $V_{(V10)}=(0.9 \pm 1.)  m/s $ \hspace{0.25cm} $Q=3.06$ \hspace{0.25cm} $wrms=2.15m/s$ }\\
\hline
\end{tabular}
\end{table}

\begin{acknowledgements}
MTS whish to thank to S\~ao Paulo State Research Foundation (FAPESP). GGS thanks to Capes for the doctoral fellowship and to Ra\'ul E. Puebla for his hints about the IDL language. We are also grateful for the comments of two anonymous referees.  
\end{acknowledgements}

\end{document}